%% file: main_text.tex
\documentclass[aps,prl,reprint,superscriptaddress,amsmath,amssymb,floatfix]{revtex4-2}

\usepackage{graphicx}
\usepackage{dcolumn}
\usepackage{bm}
\usepackage{dcolumn}
\newif\ifincludesupplement
\includesupplementtrue

\begin{document}

\title{IceCube neutrino point-source searches in the direction of the KM3NeT ultra-high-energy event}

\include{authorlist}

\begin{abstract}
    While still under construction, the KM3NeT Astroparticle Research with Cosmics in the Abyss (ARCA) detector recorded a $\sim$200 PeV neutrino on February 13th, 2023. This event is the highest-energy neutrino reported. IceCube, a cubic kilometer neutrino detector located at the geographic South Pole, has previously detected neutrinos up to approximately 10 PeV. We search for high-energy neutrinos from the location of the KM3NeT event using 15 years of IceCube data and considering three temporal hypotheses: steady or flaring in time coincidence, or at an arbitrary time. We find no evidence for neutrino emission for any of the studies performed. Correspondingly, we set upper limits on the neutrino flux from a point source in the direction of KM3-230213A. We compare these limits to KM3NeT’s estimated flux and show that an astrophysical explanation of this event is strongly constrained for a variety of spectral assumptions for a steady or transient point source with the flux inferred from the single KM3NeT ultra-high-energy event assuming a spectral index of 2.0.
\end{abstract}

\maketitle
\noindent \textit{Introduction}\label{Sec:Intro}---
    Since their discovery in 2013, the origin of high-energy astrophysical neutrinos has remained a topic of interest to the field \citep{IceCube:2013Science}. To detect high-energy neutrinos, large-volume neutrino observatories, such as the IceCube Neutrino Observatory, measure Cherenkov light produced by relativistic charged particles resulting from neutrino interactions inside the detector volume. The IceCube Neutrino Observatory is a cubic-kilometer neutrino detector embedded in the ice at the geographic South Pole. The detector consists of 5160 optical modules distributed across 86 vertical strings embedded in the Antarctic ice \citep{aartsen_icecube_2017}. 
    IceCube has made several observations in the field, such as the discovery of a high-energy astrophysical neutrino flux  \citep{IceCube:2013Science, IceCube:2014Flux} and the identification of astrophysical neutrinos associated to NGC 1068 (4.2$\sigma$) \citep{IceCube:2022Science}, TXS 0506+056 (3.5$\sigma$) \citep{IceCube:2018Science_flare}, and the Galactic plane  (4.5$\sigma$) \citep{IceCube:2023Science}. 
    KM3NeT is a neutrino telescope under construction in the Mediterranean Sea, targeting an instrumented volume of a cubic kilometer and pursuing scientific aims similar to those of IceCube. Like IceCube, it uses the Earth to shield against the atmospheric muon background, giving it a sensitive view of the Southern sky that complements IceCube's primarily Northern sky searches.
    Recently, the KM3NeT Collaboration reported the detection of a through-going muon with an estimated energy of 120 PeV. 
    KM3NeT attributed the muon to an ultra-high-energy (UHE) neutrino with a median estimated energy of 220 PeV \citep{KM3NeT:2025UHENaturePaper}. 
    This event, named KM3-230213A, has a best-fit location of (RA, DEC.) = ($94.3^\circ$, $-7.8^\circ$) with a 99\% containment radius of $3.0^\circ$. The previous highest-energy neutrino was recorded by the IceCube Collaboration with a reconstructed energy of approximately $\sim$11~PeV \citep{IceCube:2025MESECombinedDiffusePRD}. A comparison of the highest energy events reported by neutrino telescopes is provided in the End Matter. 
    
    Many investigations into the origin of the KM3NeT UHE neutrino have been conducted. Several papers published by the KM3NeT collaboration investigate nearby blazars \citep{KM3NeT:2025BlazarCounterparts, KM3NeT:2025lly}, potential galactic origin \citep{KM3NeT:2025aps}, and potential cosmogenic origin \citep{Adriani_2025} as production mechanisms for the neutrino. Additionally, many other authors and experiments have conducted their own investigations into the origin of the KM3NeT UHE neutrino. These include the KM3NeT UHE neutrino originating from a transient source \citep{Neronov:2025jfj,Yuan:2025zwe,Wang:2025lgn}, blazars \citep{Dzhatdoev:2025sdi}, cosmogenic origins \citep{Zhang:2025abk}, and primordial black holes \citep{Airoldi:2025opo,2025PhRvL.135l1003K} to name a few. Finally, investigations into the implications of the KM3NeT UHE neutrino such as a joint fit analysis between IceCube and KM3NeT \citep{KM3NeT:2025ccp}, investigations into source requirements for the origin of the neutrino \citep{Crnogorcevic:2025vou, Cermenati:2025ogl}, and searches for a diffuse UHE flux \citep{PhysRevLett.135.031001} have been performed. Notably, the search for a diffuse UHE flux reported a $2.9\sigma$ tension between IceCube and KM3NeT \citep{PhysRevLett.135.031001}. Although many source models have been proposed for the KM3NeT UHE neutrino, none are conclusive, leaving the production mechanisms uncertain. 

    In this Letter, we report the results of four searches for neutrino emission coincident with KM3-230213A using data collected by the IceCube Neutrino Observatory. These searches cover an energy range from the GeV to PeV scale, and span a variety of time-window hypotheses. In \textit{Methods} we describe the data samples and statistical techniques of our analyses. In \textit{Results} we summarize the results from each analysis. In \textit{Discussion} we summarize our conclusions and discussion of the results.

\noindent \textit{Methods}\label{Sec:Methods}---
    IceCube neutrino events can be categorized into \textit{tracks} and \textit{cascades}. \textit{Tracks} are the result of muon-neutrino ($\nu_\mu$) charged-current interactions which produce a long-lived relativistic muon and leave an elongated light pattern in the detector. \textit{Cascades} are localized, spherical signatures which result from the hadronic shower in all flavors of neutral-current interactions or from the hadronic and electromagnetic showers from electron-neutrino ($\nu_e$) and tau-neutrino ($\nu_\tau$) charged-current interactions. In this work, we perform follow-ups of KM3-230213A with IceCube data that search for steady and time-dependent emission from a point-like source in the region of the event. For the searches, we utilize a combined dataset of tracks and cascades~\citep{IceCube:SKATE} as well as the realtime track selection, known as GFU~\citep{Kintscher:2020GFU, IceCube:2023GFUICRC, IceCube:2023IceCat1}.

    This work applies an unbinned maximum-likelihood approach to four analyses: a time-integrated point-source search, a prompt transient search (centered at the time of detection by KM3NeT), a rolling flare search, and the calculation of upper limits from the non-detection of a neutrino flare by IceCube’s realtime Gamma-ray Follow Up (GFU) cluster alert system \citep{IceCube:2016RealtimeSystem, Kintscher:2020GFU}.  The likelihood construction and the associated test statistics (TS) are explained in greater detail in the Supplemental Material \citep{supplement} and \citep{Braun_2008} therein. 
    \nocite{Aartsen:2013uuv, Braun_2008, IceCube:SKATE, IceCube:2023GFUICRC,Kintscher:2020GFU, IceCube:2020FRA, IceCube:2022GWFollowup, 7yrPSTracks, IceCube:2013Science, aartsen_search_2020}
    Of these aforementioned analyses, three are archival searches using data taken during and after the KM3NeT detection time, while the fourth derives upper limits from the absence of any realtime alerts associated with the KM3NeT event. 
    The methods and analyses carried out in this paper are similar to that of \cite{aartsen_search_2020} in which IceCube carried out searches for neutrinos in the direction of the ANITA (Antarctic Impulsive Transient Antenna) experiment's neutrino candidates.

\textit{Upper Limits from Non-Detection in the Realtime System}---\label{Sec:GFUMonitoring}
    Following the announcement of the KM3NeT UHE event, we checked IceCube's realtime archive for alerts from the direction of KM3-230213A.
    IceCube's realtime platform generates several types of neutrino targets of opportunity, including the GFU-cluster alerts sent privately to imaging air Cherenkov telescopes~\citep{IceCube:2016RealtimeSystem, IceCube:2023GFUICRC, Kintscher:2020GFU, IceCube:2023IceCat1, IceCube:2020FRA, IceCube:2022GWFollowup}.
    The GFU-cluster alerts evaluate the statistical significance of any spatial and temporal clustering of events, and the pre-trial significance is used as the threshold for sending alerts (test statistic described in the Supplemental Material \cite{supplement}).
    The alert algorithm runs in two modes: all-sky, covering the regions around all incoming events with a false alarm rate (FAR) of 1 per year, and source-list, monitoring nearby, gamma-ray bright blazars with a FAR of 22 per year.

\textit{Dedicated Follow-up Searches}---\label{Time-Integrated_Search}
    All archival searches search for spatial clustering near the location of the KM3NeT UHE event. Additionally, we fit for the spectral index of the energy spectrum ($\gamma$) and the number of signal events ($n_s$), and for the rolling-flare search we additionally fit for both the flare duration and flare time. 
    In the first of the three archival searches for IceCube neutrino events in spatial coincidence with the KM3NeT UHE event presented in this work, we search for spatial clustering over the entire lifetime of IceCube data ($\sim$15 yr), assuming steady emission in the signal hypothesis.
    
    We test two different spatial probability density functions (PDFs) for the time-integrated follow-ups: a flat spatial prior and a Gaussian spatial prior. We adopt a flat spatial prior to match the methods outlined in the dedicated follow-up of the UHE event using IceCube public data performed by the KM3NeT Collaboration. KM3NeT's follow-up obtained results consistent with background expectation with a p-value of 0.07 post-trial correction \citep{KM3NeT:2025UHENaturePaper}. The Gaussian spatial prior uses the confidence radii provided in \cite{KM3NeT:2025UHENaturePaper} and is constructed as a symmetrical two-dimensional Gaussian function. Both priors use a containment radius of 99\% corresponding to a radius of $3^\circ$ originating from the best-fit location of the KM3NeT UHE event.

    In the second archival search, we consider the Gaussian spatial prior described above and search for events in boxed time windows, $\Delta t$, centered at KM3-230213A's discovery time. For this prompt transient search, we test three time windows: $10^3$, $10^5$, and $10^7$ seconds. 

    Finally, in the last archival search, we perform a rolling flare search in which we search for temporal and spatial clustering of IceCube events within the containment region of KM3-230213A. In this search we do not require that IceCube events are temporally coincident with KM3-230213A's detection time; rather, we are searching for the most significant historical Gaussian-shaped flare within the 99\% containment radius and fitting for that flare's duration, mean time, number of signal events, and spectral index.

\begin{figure}[!htbp]
    \centering
    \includegraphics[width=1\linewidth]{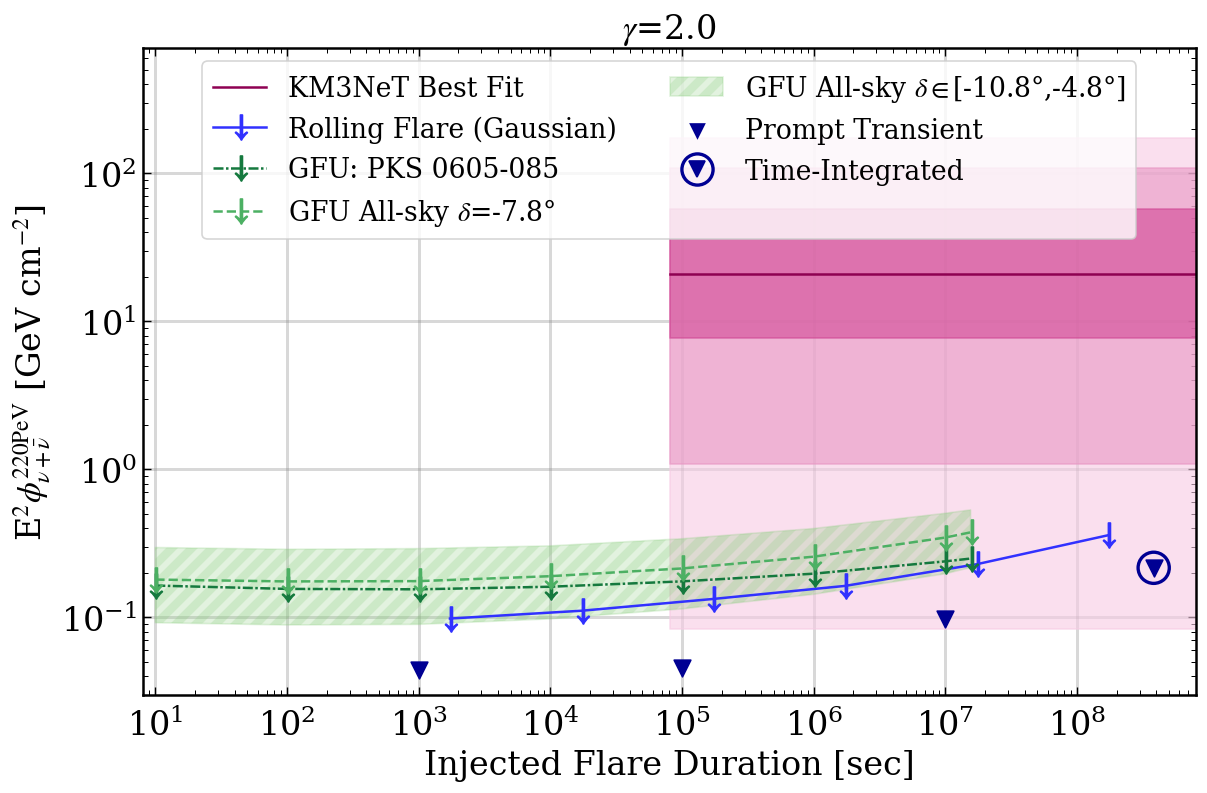}
    \includegraphics[width=1\linewidth]{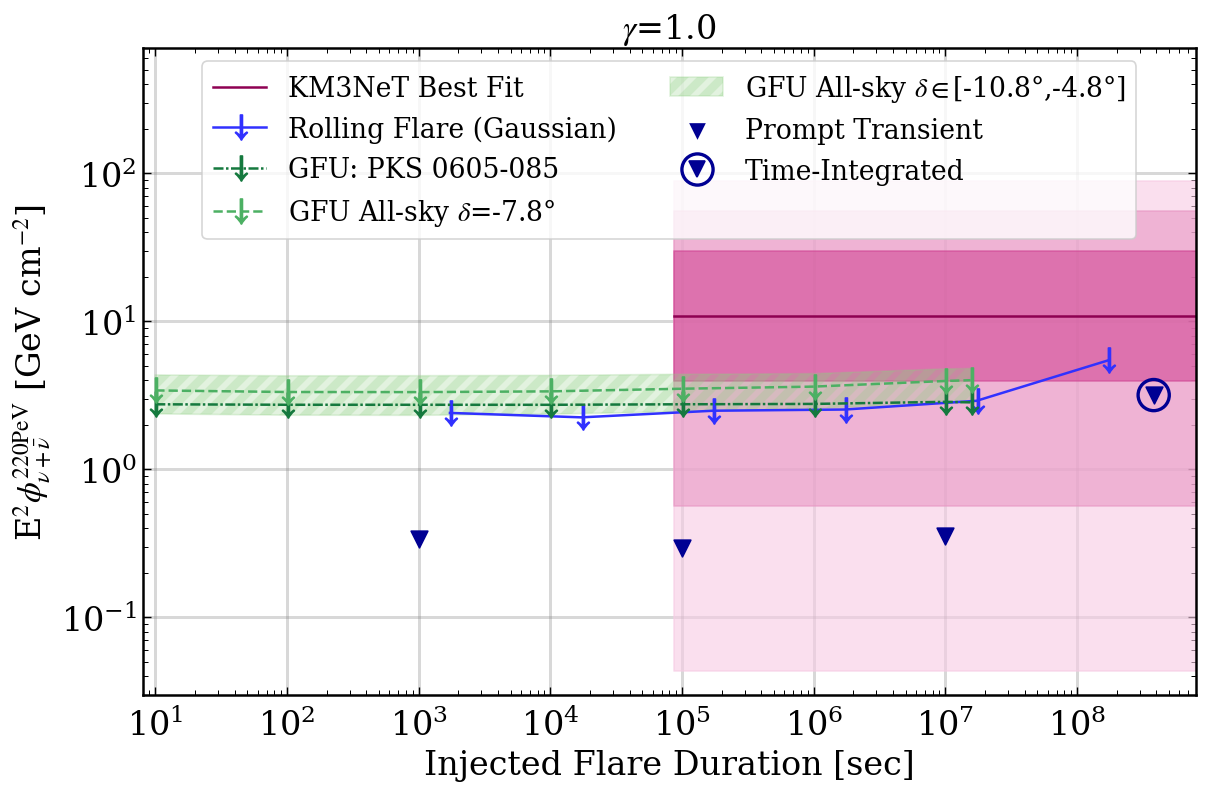}
    \caption{The 90\% confidence level upper limits on the time-integrated $\nu + \bar{\nu}$, per-flavor flux at 220 PeV from IceCube in the direction of KM3-230213A as a function of the flare duration assuming an unbroken power-law with $\gamma=2$ (top) and $\gamma=1$ (bottom). 
    The blue triangles are for the prompt transient analyses that searched time windows centered on the detection time of KM3-230213A and the time-integrated analysis (circled). 
    The solid, blue line is for the rolling flare analysis which looked for best-fit flares across the entire dataset (note: here the time-profile of the flare is treated as a Gaussian instead of a uniform boxed time-window). 
    The dashed, light-green line is for the all-sky GFU-cluster alerts both online and in the archival data for a source at $\delta$ = -7.8°, and the shaded area represents the region between -4.8° (lower edge) and -10.8° (upper edge). 
    The dashed-dotted, dark-green line is for PKS 0605-085 from the GFU-cluster source-list alerts. 
    The KM3NeT best-fit flux is indicated by the dark pink line and shaded pink regions indicating the $1\sigma$, $2\sigma$, and $3\sigma$ bands of the KM3NeT measurement and are calculated using values from \cite{KM3NeT:2025UHENaturePaper} and multiplying by the livetime of ARCA. For bottom panel, the KM3NeT flux has been recalculated according to methods outlined in \citep{supplement} to adopt a flux and error region for $\gamma = 1$. Because KM3NeT's effective area and correspondingly, its acceptance, is time-dependent due to its location, we have limited the shaded region to be for times flare durations greater than 1 day. All limits were obtained adopting the same deep inelastic scattering cross section (CSMS11 \citep{Cooper-Sarkar:2011jtt}) as KM3NeT.
    \citep{KM3NeT:2025UHENaturePaper}}
    \label{fig:ULsvTime}
\end{figure}
     
\noindent \textit{Results}---\label{Sec:Results}
No significant deviation from the background-only expectation is found in any of the analyses. All of the results are summarized in Table I and the sections hereafter.
All upper limits assume a neutrino spectrum described by a power law in energy with spectral index $\gamma$.

\textit{Upper Limits from Non-Detection in the Realtime System}---
    No all-sky GFU-cluster alerts were found near the KM3-230213A best-fit direction, neither at the event time nor in the archival alerts.
    PKS 0605-085, which was flaring in gamma-rays around the time of KM3-230213A~\citep{KM3NeT:2025BlazarCounterparts}, is monitored by the GFU-cluster algorithm, and likewise triggered no alerts at the KM3NeT event time.
    One archival alert from PKS 0605-085 was triggered by an event on February 14th, 2018 with a best-fit flare start of January 21st, 2018 and best-fit $n_s$ of 6.3 events.
    Given this non-detection by the GFU-cluster alerts, we placed 90\% confidence level upper limits on the $\nu + \bar{\nu}$ flux using the method described in the Supplemental Materials \citep{supplement}, shown as a function of the injected flare duration in Figure \ref{fig:ULsvTime}.

\begin{figure*}[!htbp]
    \centering
    \includegraphics[width=0.45\linewidth]{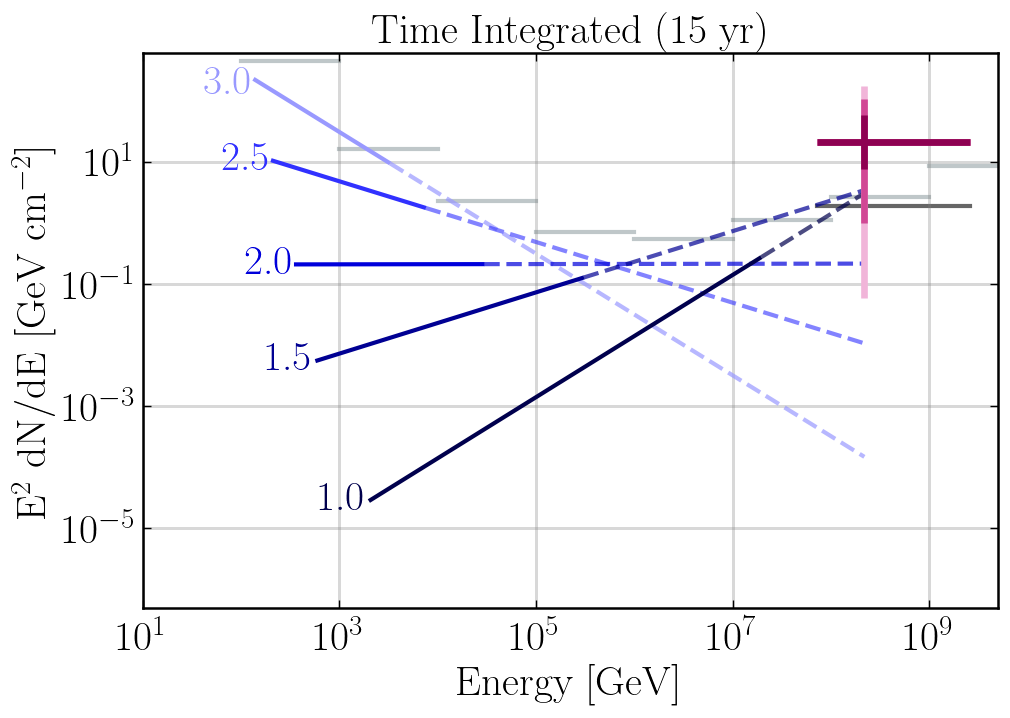}
    \includegraphics[width=0.45\linewidth]{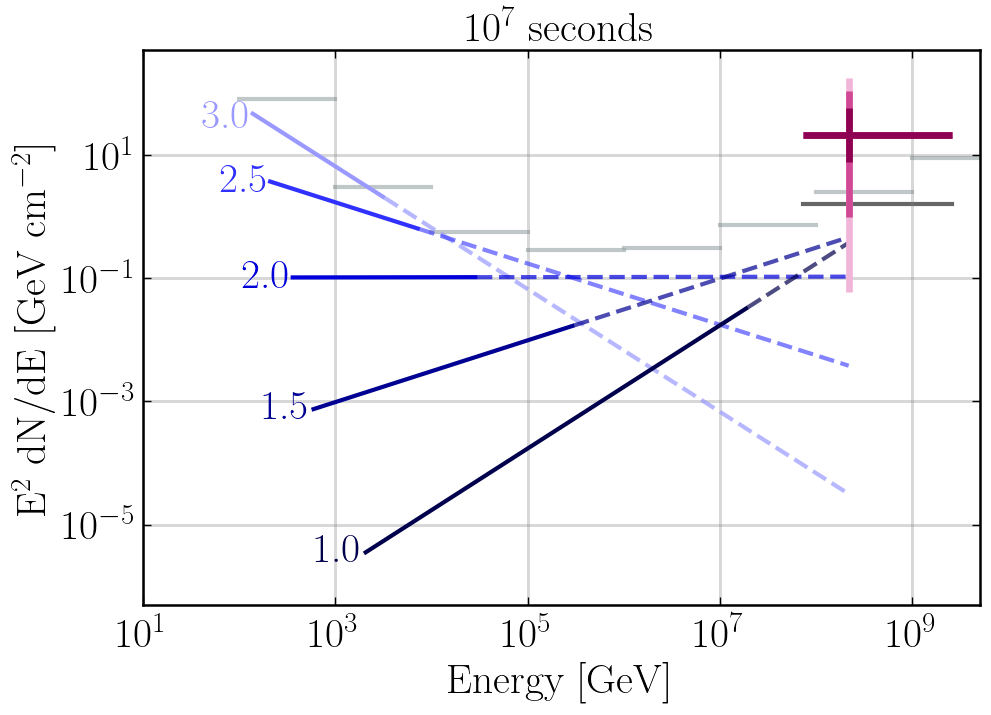}
    \includegraphics[width=0.45\linewidth]{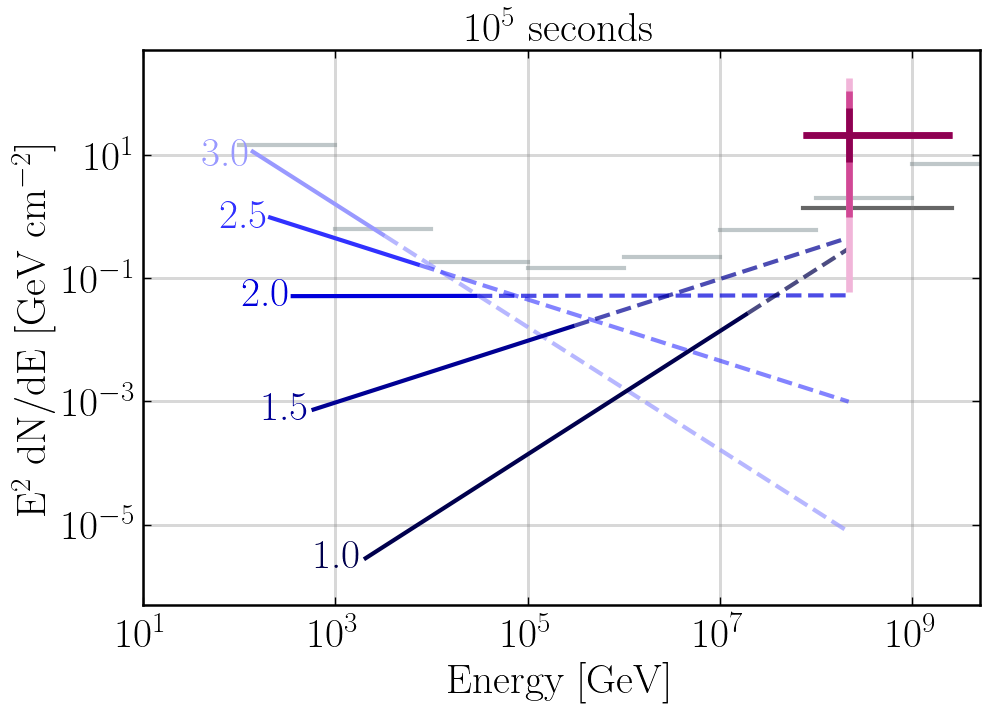}
    \includegraphics[width=0.45\linewidth]{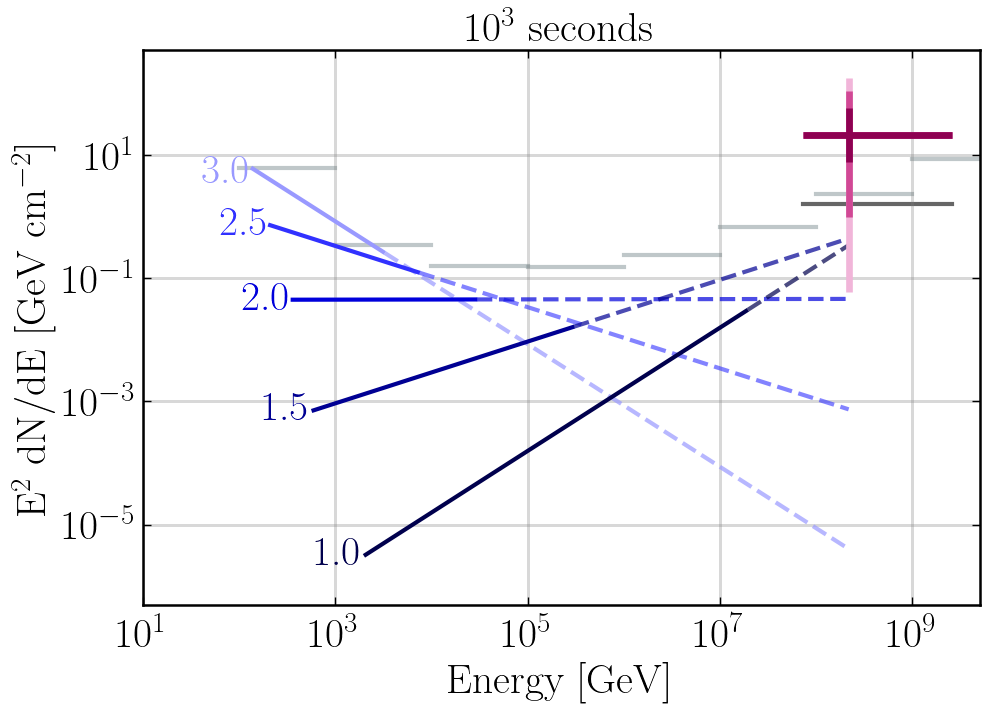}
    \caption{IceCube upper limits (90\% confidence level) and segmented (differential) limits on the time-integrated neutrino flux for a point source compatible with the localization region of KM3-230213A for several temporal search windows and various spectral indices. The blue lines are IceCube’s upper limits for $\nu + \overline{\nu}$ per-flavor flux for different spectral indices labeled next to their corresponding line. The central 90\% intervals of the expected neutrino energies for these spectra are indicated by solid lines corresponding to the central 90$\%$ energy range for a power-law with spectral index $\gamma$ at the location of the KM3NeT event. Dashed lines extend the energy range to the KM3NeT best-fit energy. Though outside the central 90\% (solid) IceCube’s most sensitive energy range, this study is sensitive to KM3-230213A energy range. Grey lines are IceCube's limits integrating over a decade of energy. The dark gray lines are integrated over the 90\% energy range from \citep{KM3NeT:2025UHENaturePaper}. All gray lines assume a spectral index of 2.0. The KM3NeT best-fit flux point is represented by a dark pink cross, with 1$\sigma$, 2$\sigma$, and 3$\sigma$ flux uncertainty indicated and calculated using values from \cite{KM3NeT:2025UHENaturePaper} and multiplying by the livetime of ARCA. }
    \label{fig:fluxes}
\end{figure*}

\textit{Time-Integrated and Prompt Transient Searches}---
We apply the unbinned maximum likelihood approach discussed in \citep{supplement} across the entire spatial prior for all of the time-integrated and prompt transient searches, with a penalty term added at each point according to the prior which represents the probability of source being at a given point. The penalty term at each point in the spatial prior is taken to be the ratio of the spatial PDF at that point over the maximum of the PDF. For further discussion, please see \citep{supplement}. We obtain a best-fit location by maximizing the test statistic across the prior, which is summarized in Table I. All results are consistent with the background expectation. In the absence of excess for the time-integrated and prompt transient searches, we set upper limits for a variety of spectral indices as depicted in Figure \ref{fig:fluxes}. 

\textit{Rolling Flare Search}---
The test statistic~\citep{supplement} is evaluated on a discrete grid of pixels spanning the 99\% containment region of the KM3NeT UHE event~\citep{KM3NeT:2025UHENaturePaper}, with an effective angular spacing of approximately $0.12^\circ$ per pixel. The grid is constructed using \texttt{healpy} \citep{healpy}, the Python \citep{python} implementation of the Hierarchical Equal Area isoLatitude Pixelization scheme (\texttt{HEALPix}; \citep{healpix}), with $N_{\rm side}=512$.

At each pixel location, a local pre-trial $p$-value is computed by comparing the observed TS to a TS distribution obtained from background-only trials. The pixel yielding the smallest pre-trial $p$-value is identified as the hottest spot. The corresponding post-trial significance is determined by comparing the hottest spot's $p$-value to a distribution of hottest spots from repeated scans of background-only realizations.

\begin{figure}[h!]
    \includegraphics[width=1\linewidth]{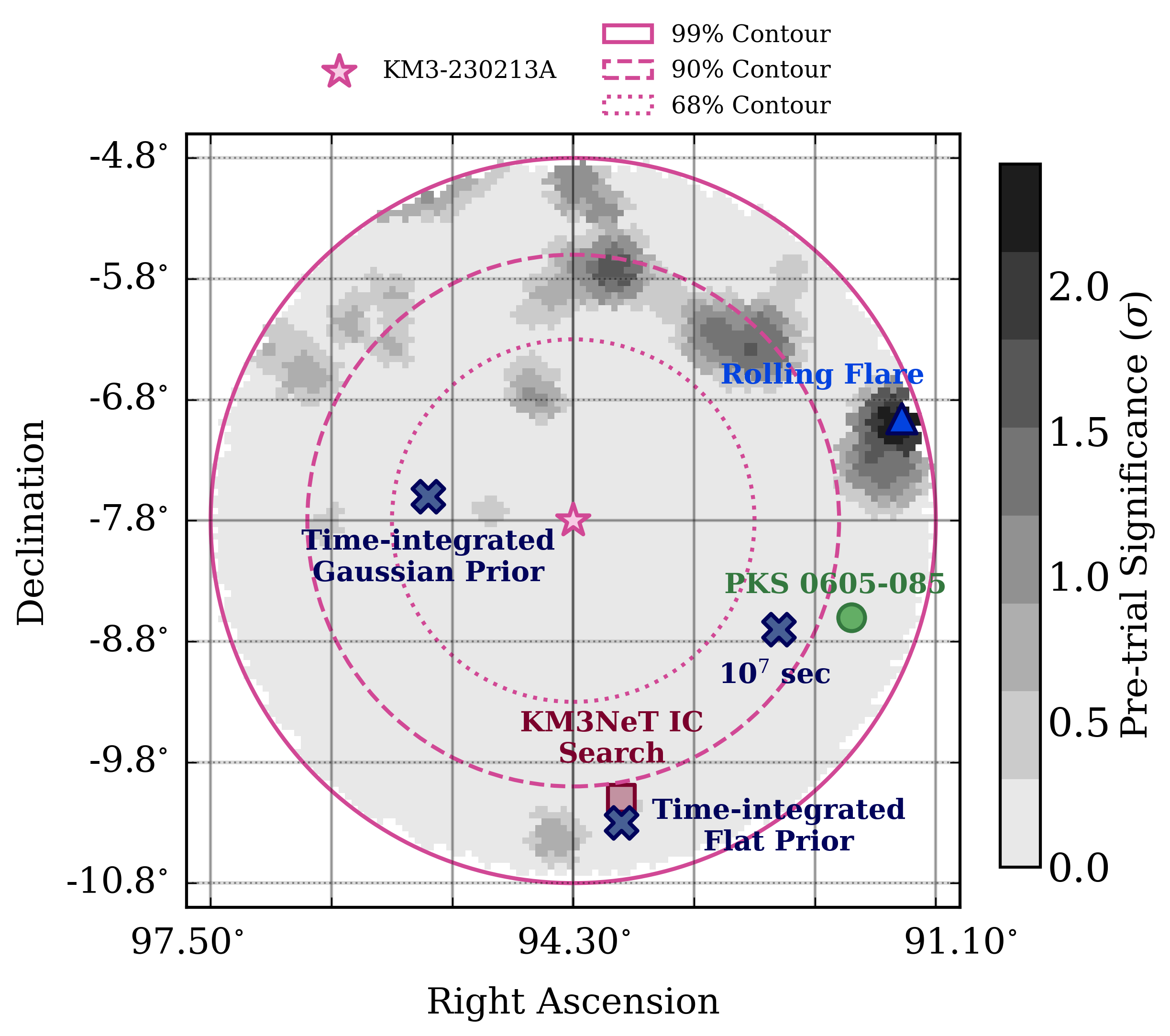}
    \caption{Pre-trial significance ($\sigma$) skymap in a $3^\circ$ radius region centered on the KM3NeT UHE event's best-fit location, shown in equatorial coordinates (J2000) for the rolling flare analysis. Pink dotted, dashed, and solid contours indicate the 68\%, 90\%, and 99\% confidence level localization regions from \cite{KM3NeT:2025UHENaturePaper}, respectively. Locations of the results in this work: the GFU monitored source PKS 0605-085 (green circle), rolling flare search (blue triangle), and time-integrated and transient searches (thick, dark blue crosses) are provided. The KM3NeT IceCube time-integrated point-source search result using public data is shown as a dark pink square. See Table I for all best-fit parameters from this work.}
    \label{fig:rolling_flare_scan}
\end{figure}

The pixel with the most significant flare in the scan region is found in equatorial coordinates (J2000) at (RA, Dec) $=(91.6^\circ, -6.96^\circ)$. The pre-trial $p$-value at this location is 0.0078. After accounting for trials across the region of the sky tested, the post-trial $p$-value at this location is 1.0, indicating that the flare is consistent with the background-only hypothesis.  The results of the rolling flare search are summarized in Table~\ref{tab:all_results} and the full scan region in local significance ($\sigma$) is shown in Figure~\ref{fig:rolling_flare_scan} (Left).

\begin{table*}[]
    \centering
    \footnotesize
    \begin{tabular}{@{}c|ccccccc}
        \hline \hline
        Analysis & \;&
        Prior &
        Best-Fit Location &
        $\hat{n}_\mathrm{s}$ &
        $\hat{\gamma}$ &
        Pre-trial  & Post-trial \\
         & & & (RA, Dec) &  &  & p-value & p-value \\\hline 
         KM3NeT Public IC Data  &  &Flat  & 93.9, -10.1   & 15.1  & \text{N/A}   & 1.6$\times 10^{-4}$ & 0.07\\  \hline
         Time-Integrated    & & Flat  & 93.9, -10.3   & 13.04     & 2.1   & 0.93 & -- \\
         Time-Integrated   & & Gaussian  & 95.5, -7.6    & 33.97     & 4.0   & 0.94 & -- \\ 
         \hline 
         Prompt Transient: 10$^3$ sec    & & Gaussian  & --     & 0.0   & --     & 1.0 & -- \\ 
         Prompt Transient: 10$^5$ sec    & & Gaussian  & --     & 0.0   & --     & 1.0 & --\\ 
         Prompt Transient: 10$^7$ sec    & & Gaussian  & 92.6, -8.7     & 3.56 & 2.30 & 0.89 & --\\  
         \hline
         Rolling Flare & & Flat & 91.6, -6.96 & 6.8 & 4.0 & 0.0078 & 1.0 \\ \hline \hline
    \end{tabular}

    \caption{Summary of Results - Results and best-fit locations for the searches performed by KM3NeT and IceCube. Note that the KM3NeT public IceCube data search does not include energy weighting and a p-value of 0.07 is consistent with background \citep{KM3NeT:2025UHENaturePaper}. For the time-dependent analyses, the time-windows reported are centered at the KM3NeT UHE detection time.}
    \label{tab:all_results}
\end{table*}

\noindent\textit{Discussion and Conclusion}\label{Sec:Discussion}---
    The recent detection of the KM3NeT UHE neutrino is of particular interest in multimessenger physics due to its high energy and potential cosmogenic origin. All results outlined in this letter are consistent with background expectation at the location of the KM3NeT UHE event. Consequently, we produce upper limits for all of our analyses - plotted in Figure \ref{fig:ULsvTime} - and obtain best-fit locations for the archival analyses which are represented in Figure \ref{fig:rolling_flare_scan}. The most constraining is the prompt transient search at short time-windows, shown for a variety of spectral indices in Figure \ref{fig:fluxes}. These results lie outside of KM3NeT's 3-sigma error and suggest that the KM3NeT UHE neutrino is not consistent with a transient source with the flux estimated in \cite{KM3NeT:2025UHENaturePaper}, assuming the spectrum extends into IceCube's energy range. Similarly, we show in our rolling flare search that archival emission is in tension at a $\sim 2\sigma$ level. Our results are consistent with the search performed using public IceCube data by KM3NeT, which obtained results consistent with background expectation \citep{KM3NeT:2025UHENaturePaper} and the results obtained in \citep{PhysRevLett.135.031001}. Future generations of neutrino detectors, such as IceCube-Gen2~\citep{IceCube_Gen2} and RNO-G (Radio Neutrino Observatory in Greenland) ~\citep{RNO-G} will improve our ability to detect these high-energy neutrinos, opening a new window into the ultra-high-energy universe.

\begin{acknowledgements}
\section{acknowledgments}
The IceCube Collaboration acknowledges the significant contributions to this manuscript from Sarah Mancina, Alicia Mand, and Riya Shah.
The authors gratefully acknowledge the support from the following agencies and institutions:
USA {\textendash} U.S. National Science Foundation-Office of Polar Programs,
U.S. National Science Foundation-Physics Division,
U.S. National Science Foundation-EPSCoR,
U.S. National Science Foundation-Office of Advanced Cyberinfrastructure,
Wisconsin Alumni Research Foundation,
Center for High Throughput Computing (CHTC) at the University of Wisconsin{\textendash}Madison,
Open Science Grid (OSG),
Partnership to Advance Throughput Computing (PATh),
Advanced Cyberinfrastructure Coordination Ecosystem: Services {\&} Support (ACCESS),
Frontera and Ranch computing project at the Texas Advanced Computing Center,
U.S. Department of Energy-National Energy Research Scientific Computing Center,
Particle astrophysics research computing center at the University of Maryland,
Michigan State University,
Astroparticle physics computational facility at Marquette University,
NVIDIA Corporation,
and Google Cloud Platform;
Belgium {\textendash} Funds for Scientific Research (FRS-FNRS and FWO),
FWO Odysseus and Big Science programmes,
and Belgian Federal Science Policy Office (Belspo);
Germany {\textendash} Bundesministerium f{\"u}r Forschung, Technologie und Raumfahrt (BMFTR),
Deutsche Forschungsgemeinschaft (DFG),
Helmholtz Alliance for Astroparticle Physics (HAP),
Initiative and Networking Fund of the Helmholtz Association,
Deutsches Elektronen Synchrotron (DESY),
and High Performance Computing cluster of the RWTH Aachen;
Sweden {\textendash} Swedish Research Council,
Swedish Polar Research Secretariat,
National Academic Infrastructure for Supercomputing in Sweden (NAISS),
and Knut and Alice Wallenberg Foundation;
European Union {\textendash} EGI Advanced Computing for research;
Australia {\textendash} Australian Research Council;
Canada {\textendash} Natural Sciences and Engineering Research Council of Canada,
Calcul Qu{\'e}bec, Compute Ontario, Canada Foundation for Innovation, WestGrid, and Digital Research Alliance of Canada;
Denmark {\textendash} Villum Fonden, Carlsberg Foundation, and European Commission;
New Zealand {\textendash} Marsden Fund;
Japan {\textendash} Japan Society for Promotion of Science (JSPS), Ministry of Education, Culture, Sports, Science and Technology (MEXT), and Institute for Global Prominent Research (IGPR) of Chiba University;
Korea {\textendash} National Research Foundation of Korea (NRF);
Switzerland {\textendash} Swiss National Science Foundation (SNSF).

\textit{Data availability}---The data that support the findings of
this Letter are not publicly available upon publication
because it is not technically feasible and/or the cost of
preparing, depositing, and hosting the data would be
prohibitive within the terms of this research project. The
data are available from the authors upon reasonable request.
\end{acknowledgements}

\bibliography{references}{}
\bibliographystyle{apsrev}

\onecolumngrid
\begin{center}
\rule{0.5\textwidth}{0.5pt}
\end{center}
\section*{End Matter}

\begin{table}[h]
 \label{tab:HighestEEvents}
    \centering
    \footnotesize
    \begin{tabular}{cc|cccc}
    \hline \hline
         Experiment & Event Morphology & Time & Deposited E & Inferred E$_{\mu}$ & Inferred E$_{\nu}$ \\
              $[$Reference$]$ & Classification & (MJD) & (PeV) & (PeV) & (PeV) \\
         \hline
         IceCube* \citep{IceCube:2025PeVCutoff} & Through-going Track & 56058.1 & - & 8.13 & - \\
         IceCube* \citep{IceCube:2023IceCat1} & Through-going Track & 56608.0 & - & - & 7.0 \\
         IceCube \citep{IceCube:2016NuMuDiffuse} & Through-going Track & 56819.2 & $2.6 \pm 0.3$ & $4.5 \pm 1.2$ & $8.7$ \\ 
         IceCube \citep{IceCube:2021Glashow} & Partially Contained Cascade & 57730.1 & 6.05 $\pm$ 0.72 & - & 6.3 \\
         IceCube \citep{IceCube:2025MESECombinedDiffusePRD} & Starting Track & 58573.3 & 4.8 &  4.3 - 9.3& 11.4$^{+2.46}_{-2.53}$ \\
         KM3NeT \citep{KM3NeT:2025UHENaturePaper} & Through-going Track & 59988.1 & - & 120$^{+110}_{-60}$ & 220$^{+570}_{-110}$ \\
    \hline \hline
    
    \end{tabular}
     \caption{All published neutrino events with a best-fit inferred energy $>$ 5 PeV.
     Events marked with an asterisk have energy estimates based on standard reconstruction algorithms rather than the resimulation-based energy inference used for the other events.
     The inferred muon energy, $E_{\mu}$ is defined as the energy of the muon as it entered the detector (or at its generation for the starting event). 
     The energy inference methods are influenced by the assumed neutrino energy spectrum which are different for each of the values reported here.} 
\end{table}
\twocolumngrid

\noindent\textit{The Highest Energy Neutrino Events}\label{Appendix:UHEEvents}-- All neutrino events detected by either IceCube or KM3NeT with a best-fit inferred energy greater than 5 PeV are listed in Table II.
All events come from different directions and are detected at different times, meaning they have no spatial or temporal correlation.
To reconstruct the energies of these events, both experiments employed an event re-simulation technique that provides a sample of Monte Carlo produced events with varied true values for the neutrino energy and systematic parameters that reproduce the observed signals in the detector~\citep{KM3NeT:2025UHENaturePaper, Chirkin:2013ResimulationReco, IceCube:2016NuMuDiffuse}.
In using the re-simulations to infer the true neutrino energy, different assumptions on the energy spectrum of astrophysical neutrinos were used for each event, which can change the range of the best-fit energy and confidence interval.

The energy estimate for KM3-230213A assumes a power-law energy spectrum with index $\gamma=-2.0$ for the simulated neutrinos.
For the IceCube track event from \cite{IceCube:2016NuMuDiffuse}, the energy was estimated assuming the best fit flux from the paper, $\gamma=-2.13$.
Similar to KM3-230213A, this track event is a ``through-going'' event, meaning that the interaction vertex occurred outside the detector and the resulting muon enters and leaves the detector without depositing all of its energy.
For through-going events, IceCube relies on the energy deposition as a function of length to estimate the muon's energy when entering the detector, which is an approximate lower bound on the neutrino energy and results in a long upper bound tail on the true neutrino energy that is dependent on the assumed energy spectrum.
The inferred neutrino energy reported in Table II is the median energy from the re-simulations \citep{IceCube:2016NuMuDiffuse}.

IceCube's Glashow event was also reconstructed using re-simulation to estimate the statistical and systematic uncertainties on the deposited energy~\citep{Chirkin:2013ResimulationReco} assuming $\gamma=-2.49$~\citep{IceCube:2021Glashow}.
In \cite{IceCube:2021Glashow}, the neutrino energy is inferred to be 6.3 PeV because of the high probability that the event was due to a Glashow resonance interaction~\citep{IceCube:2021Glashow}. 
The third event in the table, the starting event, uses the same re-simulation technique as the Glashow paper and is able to well constrain the neutrino energy due to the containment of the interaction vertex and the observation of the initial hadronic cascade along with the outgoing muon~\citep{Chirkin:2013ResimulationReco, IceCube:2025MESECombinedDiffusePRD}.
The starting event inferred energy assumes the best-fit broken power-law spectrum from \cite{IceCube:2025MESECombinedDiffusePRD}.

Two additional IceCube events have a best-fit reconstructed energy above 5 PeV and are included in Table \ref{tab:HighestEEvents}, marked with an asterisk.
These events were analyzed using the standard energy reconstructions rather than the resimulation-based technique~\citep{IceCube:2025PeVCutoff, IceCube:2023IceCat1, IceCube:2012TruncatedEnergy}.
The IceCube PeV cutoff analysis reported a best-fit muon energy of 8.13 PeV \citep{IceCube:2025PeVCutoff}.
The archival event IC131112A, included in the IceCat-1 catalog, has a best-fit muon energy of 7.0 PeV, but has a reported signalness of 0.27 because it was detected in the down-going region dominated by atmospheric muons \citep{IceCube:2023IceCat1}.
Therefore, an atmospheric-muon origin is more likely for this event than an astrophysical-neutrino origin.

Additional very-high-energy events with skimming morphologies, in which the charged particle does not enter the detector, may also be present in the existing IceCube data.
Such events are more difficult to reconstruct and distinguish from background and may not yet be included in existing event selections.

\begin{figure}[h]
        \centering
        \includegraphics[width=1\linewidth]{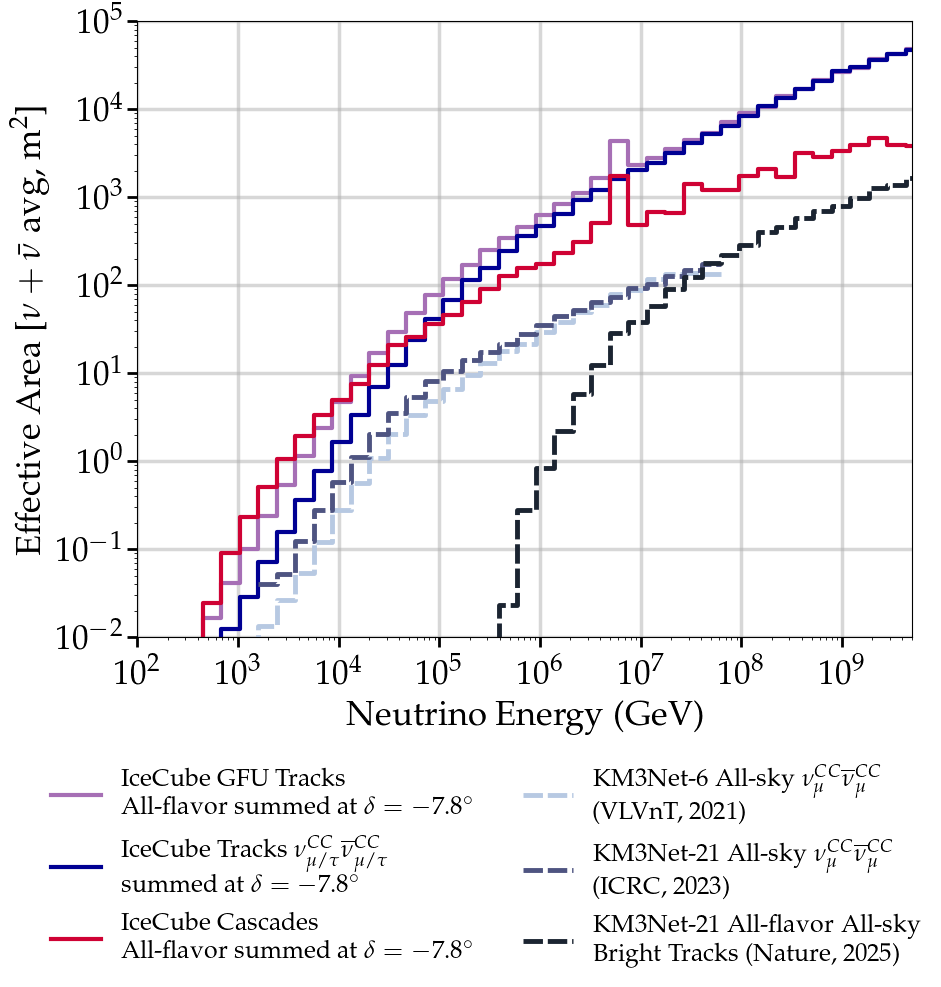}
        \caption{IceCube effective area at KM3-230213A's best-fit declination of -7.8 degrees for GFU, tracks, and cascade datasets and KM3NeT's all-sky effective area. Data for KM3NeT effective areas are taken from (in decreasing brightness) the KM3NeT Nature Paper \citep{KM3NeT:2025UHENaturePaper}, KM3NeT's 2023 ICRC proceeding \citep{muller_search_nodate}, and KM3NeT's 2021 VLVnT workshop contribution \citep{sinopoulou_atmospheric_2021}. GFU Tracks and cascades \citep{IceCube:SKATE, Kintscher:2020GFU, IceCube:2023GFUICRC} are all-flavor summed, IceCube tracks are $\nu_\mu$ and $\nu_\tau$ summed, and the three KM3NeT lines are all-sky and time-averaged. KM3NeT-6 is an upgoing track selection. Due to its position at the geographic South Pole, IceCube’s instantaneous effective area is very similar to its time-averaged effective area.}
        \label{fig:effArea}
    \end{figure}

    \noindent\textit{Data Samples and Effective Area}---
    All three archival searches use the combined track and cascade data set described in \cite{IceCube:SKATE}, whereas the GFU non-detection upper-limit calculation uses IceCube's realtime GFU event selection~\citep{Kintscher:2020GFU}. A comparison of the effective areas for these datasets and various KM3NeT datasets are seen in Figure \ref{fig:effArea}. ARCA's then-livetime consists of 335 days \citep{KM3NeT:2025UHENaturePaper}. IceCube's track datasets presented in this letter have a livetime of approximately 15 years and IceCube's cascades dataset has a livetime of approximately 12 years.

    IceCube is most sensitive to neutrinos from the northern sky (zenith $\ge85^\circ$) which must pass significant distances through the Earth to reach the detector, allowing for an effective veto of atmospheric muons, the primary source of background. At the declination of the KM3NeT neutrino, though in the southern sky (zenith $<85^\circ$), IceCube still has a large instantaneous effective area for the GFU sample and each of the datasets comprising the combined tracks and cascades dataset, as shown in Figure \ref{fig:effArea}.

\ifincludesupplement
\clearpage
\onecolumngrid
\section{Supplemental}
\section{Likelihood Methodology}\label{Appendix:LLH}

 All analyses presented in the following sections are based on a common unbinned maximum-likelihood framework for neutrino point-source searches, adapted from the method introduced in \cite{Aartsen:2013uuv}. The likelihood is written as
    \begin{equation} \label{eq:base_L}
    \mathcal{L}(n_s,\gamma ) = \prod_{j}^{M} \prod_{i \in j}^{N}\left[\frac{n_s^j}{N^j}\mathcal{S}(\textbf{x}_S, \textbf{x}_i^j, \sigma_i^j, E_i^j; \gamma) + \left(1 - \frac{n_s^j}{N^j}\right)\mathcal{B}(\sin \delta_i^j, E_i^j)\right].
    \end{equation}
This likelihood is maximized with respect to the global number of signal events $n_s$ and the assumed single, unbroken power-law spectral index $\gamma$ \citep{Braun_2008}. $N^j$ is the total number of events in the $j{\mathrm{th}}$ data sample (j represents the index) over $M$ data samples, while $n_s^j$ denotes the number of signal events per dataset and is calculated using the fraction of expected signal events from each dataset and the global number of signal events. The signal and background probability density functions are given by $\mathcal{S}$ and $\mathcal{B}$, respectively, described in detail below. The energy term of the signal PDF assumes an energy spectrum with a power-law shape, $\phi_\nu = \phi_0 \left( \frac{E_\nu}{E_0} \right)^{-\gamma}$ where $\phi_{\nu}$ is the neutrino flux, $\phi_0$ is the flux normalization at reference energy $E_0$, $E_\nu$ is the neutrino energy, and $\gamma$ is the spectral index. The reconstructed event direction is $\textbf{x}_i^j$ with angular uncertainty $\sigma_i^j$, and the source position is denoted by $\textbf{x}_S$ in right ascension $(\alpha)$ and declination $(\delta)$. The declination of the $i^{\mathrm{th}}$ event is $\delta_i^j$. As demonstrated in \cite{IceCube:SKATE}, this formulation allows for combination of tracks and cascades with differing event rates and signal purities.

The likelihood in Eq.~\ref{eq:base_L} contains both a signal PDF term, $\mathcal{S}$, and a background PDF term, $\mathcal{B}$. These PDFs are each constructed as the product of spatial and energy terms, which, together, define the probability that a given event arises from a signal source or from background. 

The spatial component of the signal PDF is modeled using either a two-dimensional Gaussian or a von Mises–Fisher distribution, depending on the event's reconstructed angular uncertainty. For events with $\sigma_i \le 7^\circ$, the small-angle approximation is valid, and the von Mises-Fisher distribution is well approximated by a 2D Gaussian:
\begin{equation} \label{S_gaussian}
\mathcal{S} = \frac{1}{2\pi \sigma_i^2}e^{-\frac{|\textbf{x}_S-\textbf{x}_i|^2}{2\sigma_i^2}} \times \mathcal{E_S}(E_i, \sin\delta_i;\gamma).
\end{equation}
For events with larger angular uncertainties ($\sigma_i > 7^\circ$), curvature of the celestial sphere becomes important and the signal spatial PDF is instead described by a von Mises–Fisher distribution:
\begin{equation} \label{S_vMF}
\mathcal{S} = \frac{1}{4\pi \sigma_i^2\sinh\left(\frac{1}{\sigma_i^2}\right)}e^{\frac{\cos(|\textbf{x}_S-\textbf{x}_i|)}{\sigma_i^2}} \times \mathcal{E_S}(E_i, \sin\delta_i;\gamma).
\end{equation}

The energy term $\mathcal{E_S}$ encodes the probability of observing an event with reconstructed energy $E_i$ at declination $\delta_i$ under a source hypothesis with a power-law spectral index $\gamma$. This term is derived from Monte Carlo (MC) simulation of signal events. The background PDF is factorized as
\begin{equation} \label{B}
\mathcal{B} = \mathcal{P_B}(\sin\delta_i) \times \mathcal{E_B}(E_i,\sin\delta_i).
\end{equation}
The spatial background term $\mathcal{P_B}$ is obtained from the observed event density as a function of declination and is normalized over the full sky. The corresponding energy term $\mathcal{E_B}$ is derived directly from data and represents the measured energy distribution of background events as a function of declination.
\subsection{GFU-cluster Alert Algorithm}
    The GFU-cluster alerts rely on the GFU event selection which runs as a filter at the South Pole and selects well-reconstructed track events.
    Events that pass the GFU filter are expressly shipped north, with a latency of around 40 seconds, and are used in several realtime analyses \citep{IceCube:2023GFUICRC, Kintscher:2020GFU, IceCube:2020FRA, IceCube:2022GWFollowup}.
    The GFU-cluster algorithm is triggered any time a new event is received and tests pixels around the event's location, referred to as the all-sky mode, and the locations of a pre-compiled list of nearby, gamma-ray bright blazars, referred to as the source-list mode \citep{IceCube:2023GFUICRC, Kintscher:2020GFU}.
    To test the neutrino flare hypothesis, the cluster alerts modify the likelihood in Eq.~\ref{eq:base_L} for a single dataset, adjusting for the time component:
    \begin{equation}\label{eq:GFU_llh}
        \mathcal{L}(n_s,\gamma,t_0 ) =  \prod_{i}^{N \in [t_0, t_1]}\left[\frac{n_s}{N}\mathcal{S}(\textbf{x}_S, \textbf{x}_i, \sigma_i, E_i; \gamma)  + \left(1 - \frac{n_s}{N}\right)\mathcal{B}(\theta_i, \phi_i, E_i)\right],
    \end{equation}
    where $t_1$ is the time of the incoming event that triggered the algorithm and the background term includes the detector zenith, $\phi$, and azimuth, $\theta$, coordinates to account for azimuthal dependence in the shorter time windows.
    The likelihood is maximized over $n_s$, $\gamma$, and the start of the flare, $t_0$, which is restricted to at most 180 days prior to the analysis-triggering event.
    The GFU-cluster test statistic (TS) includes a short time window penalty to avoid favoring the more background free region.
    The TS is given in Eq.~\ref{eqn:GFU_TS} and its construction is explained below the equation.

    The TS is converted to a p-value by comparing it to a TS distribution derived from data-driven pseudo-experiments where the GFU events are shuffled in time.
    For the all-sky mode, event clusters that result in a TS that has a 4.2$\sigma$ pre-trial significance or higher are sent as alerts.
    This threshold was selected because it results in a false alarm rate (FAR) of 1 per year.
    For the source-list mode, the pre-trial significance threshold is 3$\sigma$ resulting in a FAR of around 22 per year for all sources in the list.

    To calculate the 90\% upper limits, we performed pseudo-experiments with injected signal from simulated sources with different flare durations and spectral indices. 
    For the all-sky alerts, we also tested sources at different declinations to model the declination dependence of IceCube which is rapidly varying as a function of declination in this region of the sky.
    For each trial, we select the maximum test statistic value overlapping with the injected signal events and convert to a p-value and flux threshold where 90\% of trials result in an alert that is reported as an upper limit.

\subsection{Time Integrated Search}
For the time-integrated search, background-only trials are data-driven and generated by randomizing events in right ascension ($\alpha$). While the abundance of atmospheric background in the tracks dataset ensures that astrophysical signal contributions remain negligible when randomizing events in $\alpha$, the purity of the cascades dataset prevents the assumption that data in each declination range is background-dominated. To create background trials with this in mind, this work adopts the same signal-subtraction likelihood formalism used in previous point-source searches ~\citep{7yrPSTracks,IceCube:2023Science, IceCube:SKATE}. The signal-subtracted background expectation is: 
    \begin{equation}\label{eq:sig_sub}
    \tilde{\mathcal{D}} = \left(1-\frac{n_s^j}{N^j}\right) \mathcal{B}(\sin(\delta_i^j), E_i^j) + \frac{n_s^j}{N^j} \tilde{\mathcal{S}}(\sin(\delta_i^j), E_i^j),
    \end{equation}
    which, when substituted into Equation \ref{eq:base_L} after solving for $\left(1-\frac{n_s^j}{N^j}\right) \mathcal{B}(\sin(\delta_i^j), E_i^j)$ and including the spatial probability distribution function of the KM3NeT event, $P_K$ , yields: 
    \begin{equation}\label{eq:sigsub_L}
     \mathcal{L} (n_s, \gamma) = \prod_{j}^M \prod_{i\in j}^N \left[ \frac{n_s^j}{N} \mathcal{S} + \tilde{\mathcal{D}} (\sin(\delta_i^j), E_i^j) - \frac{n_s^j}{N^j}\tilde{\mathcal{S}} (\sin(\delta_i^j), E_i^j) \right] \cdot P_K.
    \end{equation}

    Here, the likelihood includes a data-driven background expectation consisting of $\tilde{\mathcal{D}}$, the background expectation from scrambling in $\alpha$, and $\tilde{\mathcal{S}}$, the expected contribution from the signal PDF due to residual signal from the Galactic Plane. $\mathcal{S}$ here is identical to the one in Equation \ref{eq:base_L}.
\subsection{Prompt Transient Search}
The prompt transient search modifies our signal PDF, $S$, by effectively setting the time-component of $S$ to be a uniform PDF during the on-time window and zero to all times outside of our selected time-window. Additionally, to account for low statistics in background events for short time-windows we introduce a Poisson term to our likelihood described in Eq.~\ref{eq:base_L}, similar to the methods described in \cite{aartsen_search_2020}:
    \begin{equation}
        \lambda = \frac{(n_s + n_b)^N}{N!}\cdot e^{-(n_s + n_b)},
    \end{equation}
    where $n_s$ and $n_b$ are the expected number of  observed signal and background events, respectively, and $N$ is the number of events in the on-time window. This gives the corresponding likelihood: 
    \begin{equation}\label{eq:transient_L}
        \mathcal{L} = \prod_j^M \frac{(n_s^j + n_b^j)^{N^j}}{N^j!}\cdot e^{-(n_s + n_b)}\prod_{i \in j}^N \left[ \frac{n_s^j}{N^j} \mathcal{S} + \left(1-\frac{n_s^j}{N^j}\right)\mathcal{B}\right] \cdot P_K.
    \end{equation}
    
\subsection{Rolling Flare Search}\label{Rolling_Flare_Search}
This work includes a time-dependent search for neutrino emission from locations within the 99\% localization contour of KM3-230213A. The method follows the unbinned, time-dependent maximum-likelihood framework of \cite{Braun_2008} and is based on a modified form of Eq.~\ref{eq:base_L}. The likelihood is given by:
\begin{equation} \label{eq:TD_likelihood}
\mathcal{L}(n_s,\gamma, T_0, \sigma_t) = \prod_{j}^{M} \prod_{i \in j}^{N}\frac{n_s^j}{N^j}\mathcal{S} \times \mathcal{T_S}(T_0, \sigma_t)+ (1 - \frac{n_s^j}{N^j})\mathcal{B}\times \mathcal{T_B}.
\end{equation}
The $\mathcal{S}$ and $\mathcal{B}$ are identical to those defined in Eq.~\ref{eq:base_L}, with the addition of time-dependent signal, $\mathcal{T}_\mathcal{S}$, and background, $\mathcal{T}_\mathcal{B}$, terms (See below). This search fits for the number of signal events, $n_s$, the spectral index, $\gamma$, the mean time of the flare, $T_0$, and the Gaussian half-width duration of the flare, $\sigma_t$. 
The maximum allowed flare half-width for this search is restricted to 289 days, corresponding to the time between  the KM3-230213A detection and the end of the combined track and cascade data set.

The rolling flare analysis does not require the application of the signal-subtraction likelihood shown in Eq.~\ref{eq:sigsub_L}. In this time-dependent framework, background trials are constructed by randomizing both the event arrival times and $\alpha$, which suppresses any residual spatial or temporal clustering from true signal. As a result, the background-only hypothesis is accurately represented without explicit signal removal.

The time-dependent signal and background terms are added to Eq.~\ref{eq:base_L} to produce Eq.~\ref{eq:TD_likelihood}. The signal time PDF, $\mathcal{T}_\mathcal{S}$, is modeled as a Gaussian,
\begin{equation} \label{eq:TD_S}
\mathcal{T_S}(T_0, \sigma_t) = \frac{1}{\sqrt{2\pi\sigma_t^2}}e^{-\frac{(t_i-T_0)^2}{2\sigma_t^2}},
\end{equation}
where $T_0$ is the central time of the flare, $\sigma_t$ is the Gaussian half-width of the flare, and $t_i$ is the time of event $i$. The background time PDF, $\mathcal{T}_\mathcal{B}$, is approximated to be uniform, $\mathcal{T_B} = 1/T_{\rm tot}$, where $T_{\rm tot}$ is the total livetime of the sample.

\section{Test Statistic Construction}\label{Appendix:TS}
The GFU-cluster alerts define the test statistic as:
\begin{equation}\label{eqn:GFU_TS}
    TS = -2 \log \left[ \frac{\mathcal{L}(n_s = 0, \hat{t}_0, t_1)}{\mathcal{L}(\hat{n}_s, \hat{\gamma}, \hat{t}_0, t_1)} \times \frac{\mathcal{U}(t_1 - T_{\mathrm{max}}, t_1)}{\mathcal{U}(\hat{t}_0, t_1)} \right],
\end{equation}
where $\mathcal{U}$ is the total good-detector uptime between two defined times and $T_{\mathrm{max}}$ is the maximum allowed time window duration (180 days). The uptime term adds a penalty which discourages fitting shorter time windows where there is less background.

For both of the time-integrated searches we obtain a test statistic of: 
    \begin{equation}
    \rm TS = \text{max}\left(\sum_{i=1}^N2\log\left[ \frac{\mathcal{L}(\hat{n}_s, \hat{\gamma})}{\mathcal{L}(n_s=0)}\right] + 2\log\left[\frac{P_K(\mathbf{x}_s)}{P_K(\mathbf{x}_0)}\right]\right).
    \end{equation}

For the transient search, we derive the test statistic to be: 
    \begin{equation}
        \mathrm{TS} = \mathrm{max} \left(-2n_s + \sum_{i=1}^{N}2\log\left[1 + \frac{n_s\cdot S_i}{n_b\cdot B_i}\right] + 2\log\frac{P_K\left(\mathbf{x_s}\right)}{P_K\left(\mathbf{x_0}\right)}\right).
    \end{equation}

Finally, for the rolling flare search, the test statistic is defined as
\begin{equation} \label{eq:TD_TS}
\mathrm{TS} = -2 \log\left[\frac{T_{\rm tot}}{\hat{\sigma}_t}\times \frac{\mathcal{L}(n_s=0)}{\mathcal{L}(\hat{n}_s, \hat{\gamma}, \hat{\sigma}_t, \hat{T}_0)}\right],
\end{equation}
where the factor $T_{\rm tot}/\hat{\sigma}_t$ accounts for the larger effective number of trials associated with shorter-duration flares, which are more numerous within the total observation time and therefore have a higher probability of producing a background fluctuation. 

\section{Constructing KM3NeT Flux for $\gamma = 1$}\label{Appendix:fluxCalc}
We calculate flux according to:
\begin{equation}
    \Phi(E) = \hat\phi \cdot \left[ \frac{E}{GeV}\right]^{-\gamma},
\end{equation}
where $\Phi (E)$ is the flux as a function of energy with units of $\left[\frac{1}{\mathrm{GeV}\cdot \mathrm{sr} \cdot \mathrm{sec}\cdot \mathrm{cm}^2}\right]$, $\hat\phi$ is the flux normalization with units of $\left[\frac{\mathrm{GeV}^{\gamma -1 }}{\mathrm{sr} \cdot \mathrm{sec}\cdot \mathrm{cm}^2}\right]$, and $E$ is energy in units of GeV. We may obtain the flux normalization with methods similar to those described in \citep{KM3NeT:2025UHENaturePaper} using the following equation: 
\begin{equation}
    n_{exp} = \hat\phi \int_{E_{MIN}}^{E_{MAX}} 4\pi \cdot\Delta T\cdot A(E) \cdot \left[ \frac{E}{GeV}\right]^{-\gamma}dE.
\end{equation}

Here, $n_{exp}$ is the number of expected neutrinos for a per-flavor $\nu + \bar{\nu}$ all-sky flux which for the purposes of our calculation is 1 for KM3NeT's single observed event. The integration bounds $E_{MIN}, E_{MAX}$ are set to the central 90\% neutrino energy range reported by KM3NeT. $\Delta T$ is the livetime of KM3NeT at the time of detection (335 days), and $A(E)$ is KM3NeT's sky-averaged effective area \citep{KM3NeT:2025UHENaturePaper}. Finally, the $4\pi$ term accounts for the fact that the effective area is sky-averaged. After substituting 1.0 in for $\gamma$, and solving for the flux normalization for one observed event, we obtain our flux. Finally, we multiply by the livetime of ARCA and $4\pi$ in order to obtain the time-integrated flux for our calculations.


\end{document}

%% file: authorlist.tex
\affiliation{III. Physikalisches Institut, RWTH Aachen University, D-52056 Aachen, Germany}
\affiliation{Department of Physics, University of Adelaide, Adelaide, 5005, Australia}
\affiliation{Dept. of Physics and Astronomy, University of Alaska Anchorage, 3211 Providence Dr., Anchorage, AK 99508, USA}
\affiliation{School of Physics and Center for Relativistic Astrophysics, Georgia Institute of Technology, Atlanta, GA 30332, USA}
\affiliation{Dept. of Physics, Southern University, Baton Rouge, LA 70813, USA}
\affiliation{Dept. of Physics, University of California, Berkeley, CA 94720, USA}
\affiliation{Lawrence Berkeley National Laboratory, Berkeley, CA 94720, USA}
\affiliation{Institut f{\"u}r Physik, Humboldt-Universit{\"a}t zu Berlin, D-12489 Berlin, Germany}
\affiliation{Fakult{\"a}t f{\"u}r Physik {\&} Astronomie, Ruhr-Universit{\"a}t Bochum, D-44780 Bochum, Germany}
\affiliation{Universit{\'e} Libre de Bruxelles, Science Faculty CP230, B-1050 Brussels, Belgium}
\affiliation{Vrije Universiteit Brussel (VUB), Dienst ELEM, B-1050 Brussels, Belgium}
\affiliation{Dept. of Physics, Simon Fraser University, Burnaby, BC V5A 1S6, Canada}
\affiliation{Department of Physics and Laboratory for Particle Physics and Cosmology, Harvard University, Cambridge, MA 02138, USA}
\affiliation{Dept. of Physics, Massachusetts Institute of Technology, Cambridge, MA 02139, USA}
\affiliation{Dept. of Physics and The International Center for Hadron Astrophysics, Chiba University, Chiba 263-8522, Japan}
\affiliation{Department of Physics, Loyola University Chicago, Chicago, IL 60660, USA}
\affiliation{Dept. of Physics and Astronomy, University of Canterbury, Private Bag 4800, Christchurch, New Zealand}
\affiliation{Dept. of Physics, University of Maryland, College Park, MD 20742, USA}
\affiliation{Dept. of Astronomy, Ohio State University, Columbus, OH 43210, USA}
\affiliation{Dept. of Physics and Center for Cosmology and Astro-Particle Physics, Ohio State University, Columbus, OH 43210, USA}
\affiliation{Niels Bohr Institute, University of Copenhagen, DK-2100 Copenhagen, Denmark}
\affiliation{Dept. of Physics, TU Dortmund University, D-44221 Dortmund, Germany}
\affiliation{Dept. of Physics and Astronomy, Michigan State University, East Lansing, MI 48824, USA}
\affiliation{Dept. of Physics, University of Alberta, Edmonton, Alberta, T6G 2E1, Canada}
\affiliation{Erlangen Centre for Astroparticle Physics, Friedrich-Alexander-Universit{\"a}t Erlangen-N{\"u}rnberg, D-91058 Erlangen, Germany}
\affiliation{Physik-department, Technische Universit{\"a}t M{\"u}nchen, D-85748 Garching, Germany}
\affiliation{D{\'e}partement de physique nucl{\'e}aire et corpusculaire, Universit{\'e} de Gen{\`e}ve, CH-1211 Gen{\`e}ve, Switzerland}
\affiliation{Dept. of Physics and Astronomy, University of Gent, B-9000 Gent, Belgium}
\affiliation{Dept. of Physics and Astronomy, University of California, Irvine, CA 92697, USA}
\affiliation{Karlsruhe Institute of Technology, Institute for Astroparticle Physics, D-76021 Karlsruhe, Germany}
\affiliation{Karlsruhe Institute of Technology, Institute of Experimental Particle Physics, D-76021 Karlsruhe, Germany}
\affiliation{Dept. of Physics, Engineering Physics, and Astronomy, Queen's University, Kingston, ON K7L 3N6, Canada}
\affiliation{Department of Physics {\&} Astronomy, University of Nevada, Las Vegas, NV 89154, USA}
\affiliation{Nevada Center for Astrophysics, University of Nevada, Las Vegas, NV 89154, USA}
\affiliation{Dept. of Physics and Astronomy, University of Kansas, Lawrence, KS 66045, USA}
\affiliation{UCLouvain, Centre for Cosmology, Particle Physics and Phenomenology, CP3, Chemin du Cyclotron 2, 1348 Louvain-la-Neuve, Belgium}
\affiliation{Department of Physics, Mercer University, Macon, GA 31207-0001, USA}
\affiliation{Dept. of Astronomy, University of Wisconsin{\textemdash}Madison, Madison, WI 53706, USA}
\affiliation{Dept. of Physics and Wisconsin IceCube Particle Astrophysics Center, University of Wisconsin{\textemdash}Madison, Madison, WI 53706, USA}
\affiliation{Institute of Physics, University of Mainz, Staudinger Weg 7, D-55099 Mainz, Germany}
\affiliation{Department of Physics, Marquette University, Milwaukee, WI 53201, USA}
\affiliation{Institut f{\"u}r Kernphysik, Universit{\"a}t M{\"u}nster, D-48149 M{\"u}nster, Germany}
\affiliation{Bartol Research Institute and Dept. of Physics and Astronomy, University of Delaware, Newark, DE 19716, USA}
\affiliation{Dept. of Physics, Yale University, New Haven, CT 06520, USA}
\affiliation{Columbia Astrophysics and Nevis Laboratories, Columbia University, New York, NY 10027, USA}
\affiliation{Dept. of Physics, University of Oxford, Parks Road, Oxford OX1 3PU, United Kingdom}
\affiliation{Dipartimento di Fisica e Astronomia Galileo Galilei, Universit{\`a} Degli Studi di Padova, I-35122 Padova PD, Italy}
\affiliation{Dept. of Physics, Drexel University, 3141 Chestnut Street, Philadelphia, PA 19104, USA}
\affiliation{Physics Department, South Dakota School of Mines and Technology, Rapid City, SD 57701, USA}
\affiliation{Dept. of Physics, University of Wisconsin, River Falls, WI 54022, USA}
\affiliation{Dept. of Physics and Astronomy, University of Rochester, Rochester, NY 14627, USA}
\affiliation{Department of Physics and Astronomy, University of Utah, Salt Lake City, UT 84112, USA}
\affiliation{Dept. of Physics, Chung-Ang University, Seoul 06974, Republic of Korea}
\affiliation{Oskar Klein Centre and Dept. of Physics, Stockholm University, SE-10691 Stockholm, Sweden}
\affiliation{Dept. of Physics and Astronomy, Stony Brook University, Stony Brook, NY 11794-3800, USA}
\affiliation{Dept. of Physics, Sungkyunkwan University, Suwon 16419, Republic of Korea}
\affiliation{Institute of Physics, Academia Sinica, Taipei, 11529, Taiwan}
\affiliation{Dept. of Physics and Astronomy, University of Alabama, Tuscaloosa, AL 35487, USA}
\affiliation{Dept. of Astronomy and Astrophysics, Pennsylvania State University, University Park, PA 16802, USA}
\affiliation{Dept. of Physics, Pennsylvania State University, University Park, PA 16802, USA}
\affiliation{Dept. of Physics and Astronomy, Uppsala University, Box 516, SE-75120 Uppsala, Sweden}
\affiliation{Dept. of Physics, University of Wuppertal, D-42119 Wuppertal, Germany}
\affiliation{Deutsches Elektronen-Synchrotron DESY, Platanenallee 6, D-15738 Zeuthen, Germany}

\author{R. Abbasi}
\affiliation{Department of Physics, Loyola University Chicago, Chicago, IL 60660, USA}
\author{M. Ackermann}
\affiliation{Deutsches Elektronen-Synchrotron DESY, Platanenallee 6, D-15738 Zeuthen, Germany}
\author{J. Adams}
\affiliation{Dept. of Physics and Astronomy, University of Canterbury, Private Bag 4800, Christchurch, New Zealand}
\author{J. A. Aguilar}
\affiliation{Universit{\'e} Libre de Bruxelles, Science Faculty CP230, B-1050 Brussels, Belgium}
\author{M. Ahlers}
\affiliation{Niels Bohr Institute, University of Copenhagen, DK-2100 Copenhagen, Denmark}
\author{J.M. Alameddine}
\affiliation{Dept. of Physics, TU Dortmund University, D-44221 Dortmund, Germany}
\author{S. Ali}
\affiliation{Dept. of Physics and Astronomy, University of Kansas, Lawrence, KS 66045, USA}
\author{N. M. Amin}
\affiliation{Bartol Research Institute and Dept. of Physics and Astronomy, University of Delaware, Newark, DE 19716, USA}
\author{K. Andeen}
\affiliation{Department of Physics, Marquette University, Milwaukee, WI 53201, USA}
\author{C. Arg{\"u}elles}
\affiliation{Department of Physics and Laboratory for Particle Physics and Cosmology, Harvard University, Cambridge, MA 02138, USA}
% ...
\date{\today}